\documentclass[superscriptaddress,twocolumn,showpacs,preprintnumbers,amsmath,amssymb]{revtex4}

\usepackage{graphicx}
\usepackage{float}
\usepackage[latin1]{inputenc}
\usepackage{times}
\usepackage{makeidx}
\usepackage{bm}
\usepackage{amsfonts}
\usepackage{fancybox}
\usepackage{amssymb}
\usepackage[isu,scriptsize]{caption}


\begin{document}

\title{\textit{First principles} Kinetic-Collective thermal conductivity of semiconductors}

\author{P. Torres}
\author{A. Torello}
\author{J. Bafaluy}
\author{J. Camacho}
\affiliation{Departament de F\'isica, Universitat Aut\`onoma de
Barcelona, 08193 Bellaterra, Catalonia, Spain}
\author{X. Cartoix\`a}
\affiliation{Departament d'Enginyeria Electr\`onica, Universitat Aut\`onoma de
Barcelona, 08193 Bellaterra, Catalonia, Spain}
\author{F. X. Alvarez}
\email{xavier.alvarez@uab.es}
\affiliation{Departament de F\'isica, Universitat Aut\`onoma de
Barcelona, 08193 Bellaterra, Catalonia, Spain}

\date{\today}
\begin{abstract}
A fully predictive Kinetic Collective Model using {\it first principles} phonon spectra and relaxation times is presented. Thermal conductivity values obtained for Si, Ge, C (diamond) and GaAs in a wide range of sizes and temperatures have good agreement with experimental data without the use of any fitting parameter. This validation of the model open the door to discuss how the precise combination of kinetic and collective contributions to heat transport could provide a useful framework to interpret recent complex experiments displaying non-Fourier behavior. 
\end{abstract}
\maketitle

Recent experiments in thermal transport since the appearance of ultra-fast laser techniques, measuring the effective thermal conductivity using heaters with different sizes and working in different excitation frequency ranges, have shown that the Fourier law breaks down at reduced size and time scales~\cite{Siemens2010,Minnich2011,Regner2013,Johnson2013,Wilson2014,Hu2015,Hoogeboom-Pot2015}. To understand the origin of this new behaviour, the authors try to obtain the thermal conductivity spectral distribution (TCSD) in terms of the phonon relaxation times or the phonon mean free paths (MFP)~\cite{Minnich2011,Regner2013,Johnson2013,Wilson2014,Hu2015,Hoogeboom-Pot2015,Mohammed2015}. 

To extract the TCSD from experiments, a microscopic insight is needed. In the standard kinetic framework, thermal conductivity is obtained by simply adding independent single mode contributions~\cite{Hu2015}. This approach is known to be valid for highly resistive materials at large size scales. However, it is widely accepted that although normal (N) scattering does not contribute to the thermal resistivity, it can cause qualitative differences in heat flow~\cite{Peierls2001,Ziman2001}: momentum conservation does not allow a rapid relaxation of thermal disturbances and heat flux can change to a regime where phonons are highly correlated (collective regime)~\cite{DeTomas2014}. In this case the contribution of the participating modes to the total thermal transport can change dramatically~\cite{DeTomas2015}. Several works have focused on obtaining a proper framework to address the effect of normal scattering, either iteratively solving the Boltzmann Transport Equation (BTE)~\cite{Broido2007, Lee2015} or keeping the kinetic description and changing to a different quasiparticle (relaxon) \cite{Cepellotti2016}. So far neither of these models have given a definitive picture to interpret ultra-fast heating experiments.

The Kinetic-Collective Model (KCM) is derived from the solution to the Boltzmann Transport Equation (BTE) \cite{DeTomas2014} expanded in terms of eigenstates of the normal collision operator~\cite{Guyer1966a}. Thus it is a natural framework to understand and analyse systems where phenomena related to momentum conservation are expected to be important, such as graphene~\cite{Lee2015,Cepellotti2015} or group IV materials~\cite{DeTomas2014a}.



The purpose of this work is to show that the KCM provides a useful framework to describe heat transport at all time and length scales. On one hand, KCM in combination with {\it first principles} calculations of microscopic magnitudes is able to predict the thermal conductivity of a wide range of semiconductors both in bulk and nanoscale samples without using any fitting parameters. These experiments are re-interpreted at the light of the framework providing a physical insight of the behaviour of thermal conduction as size and temperature change. On the other hand, we show that the differences between the kinetic and collective contributions are key to interpret recent results in ultra-fast heating experiments. 

In the KCM it is possible to split the thermal conductivity into a kinetic and a collective contribution weighed through the use of a switching factor $\Sigma \in [0,1]$, measuring the relative importance of the normal and resistive scattering rates \citep{Guyer1966,DeTomas2014}. The resistive terms in each contribution are treated in a different way. While the boundary scattering is included trough the Mathiessen rule in the kinetic contribution, a form factor $F$, determined by the sample geometry alone through an effective length $L_{\rm eff}$, includes the size effects on the collective term. Thus, in the calculation of $\tau_{c}$ (collective mean free time) the umklapp and impurity scattering are the only processes considered. In bulk materials $F=1$ and the equation depends only on intrinsic scattering events. Notice that normal scattering rates do not contribute on any of both conduction regimes but just in the switching factor. The equation for thermal conductivity is

\begin{equation}
\kappa=\kappa_{k}+\kappa_{c} , \label{KCM}
\end{equation}

where

\begin{equation}
\kappa_{k}(T)=(1-\Sigma(T))\int \hbar\omega\frac{\partial n}{\partial T} v^2(\omega)\tau_{k}(\omega)DOS(\omega)d\omega
\end{equation}
\begin{equation}
\kappa_{c}(T)=\Sigma(T) F (L_{\rm eff}) \int \hbar\omega\frac{\partial n}{\partial T} v^2(\omega) \tau_{c}(T)DOS(\omega)d\omega
\end{equation}
\begin{equation}
\Sigma(T)=\frac{1}{1+\frac{\tau_{n}(T)}{\tau_{R}(T)}} \label{eq_sigma}
\end{equation}

In the frequency domain an important difference is that while the kinetic mean free time $\tau_{k_{i}}$ is different for all the phonons, the collective mean free time $\tau_c$ is the same for all of them. In this framework not all the energy is carried kinetically by independent collisions as thought, but part of this energy is carried by the called collective phonons. A complete description of the model and explanation of the different contributions can be obtained elsewhere~\cite{DeTomas2014,DeTomas2014a}.

We calculate all the needed magnitudes in Eq.~(\ref{KCM}) from {\it first principles} using the {\sc Quantum ESPRESSO} package~\cite{Giannozzi2009}, which implements Density Functional Theory (DFT)~\cite{Hohenberg1964,Kohn1965} under the Local Density Approximation in the parametrization of Perdew and Zunger~\cite{Perdew1981}. Core electrons were accounted for with norm-conserving pseudo-potentials of the Von Barth-Car type~\cite{VonBarthU.andCar,DalCorso1993}. Plane waves were cut off at an energy of 60 Hartree and Born effective charges and dielectric tensor were employed for GaAs to account for its polar behavior. Finally, small atomic cartesian displacements in a 3x3x3 super-cell with 216 atoms up to 3rd neighbours were performed to compute second and third order force constants. A 20x20x20 q-point grid is used for phonon Brillouin zone sampling in such calculations, while a 160x160x160 mesh is used for the density of states (DOS) calculations. Normal and umklapp phonon relaxation times (i.e. mean free times) are obtained through the anharmonic force constants. For this we use the open code package ALAMODE \cite{Tadano2014}, where splitting of normal and umklapp events have been manually implemented in the code. Extrapolation of the latter values have been done for low frequencies in the DOS mesh sampling.

\begin{figure}
\includegraphics[width=\columnwidth]{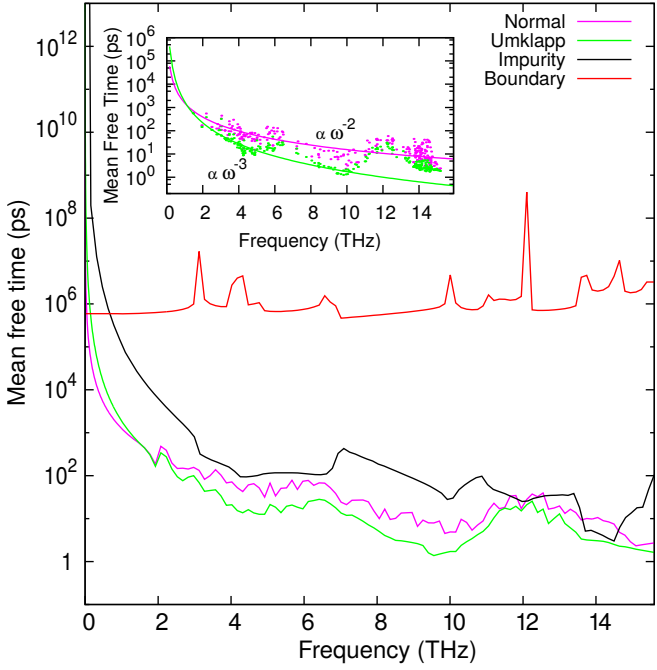}
\caption{Impurity, boundary, normal and umklapp relaxation times for bulk silicon in terms of frequency at T=300K. Impurity is obtained with Eq. (\ref{eg:mft_impurity}), boundary curve following Ref. \cite{Zhang2007} and umklapp and normal are calculated averaging the \textit{ab-initio} results over bins. Rough \textit{ab-initio} data are plotted in the inset together with the analytic approximations ($\omega^{-2}$ for normal and $\omega^{-3}$ for umklapp).} \label{fig_bulk_taus_w}
\end{figure}

For the boundary collision rates in the kinetic regime we use Cassimir expression \cite{Casimir} $ \tau_{b}(\omega)=L_{eff}/v(\omega)$, where $L_{eff}$ is the wire diameter, $1.12l$ for rods and $2.25h$ for thin films \cite{DeTomas2014}. Geometry effects in small samples and the effect of roughness has been demonstrated to be important for the thinnest nanowires \cite{Asheghi1997,Zhang2007,Martin2009b}. 

For the impurity collision rate, we use Tamura's expression~\cite{Capinski1999}
\begin{equation}
\tau_{I}^ {-1} = \frac{\pi}{6} \Gamma D_{\omega} \omega^2\label{eg:mft_impurity}
\end{equation}
where $D_{\omega}$ is the density of states and $\Gamma$ is the mass variance of the sample depending on the isotopic abundance of the sample. Notice that all these magnitudes are calculated and no free parameters are used. Fig. \ref{fig_bulk_taus_w} represents the obtained frequency dependent phonon relaxation times for all the considered scattering mechanisms in a 3~mm bulk Si sample (the normal and umklapp scattering curves correspond to binning of the points from the inset in Fig.~\ref{fig_bulk_taus_w}). For low frequencies we use theoretically derived expressions. We use Han's expressions \cite{Han} for the umklapp processes, providing a smooth transition from $ \omega^3 $ to $ \omega^2 $. For normal processes, Herring's expression \cite{Herring1954} where $ \tau_N $ $ \alpha$ $ \omega^2T^3 $ is used. Using analitical expressions for low frequencies, accurate results can be achieved even with a coarse k-point grid. 

\begin{figure}
\includegraphics[width=\columnwidth]{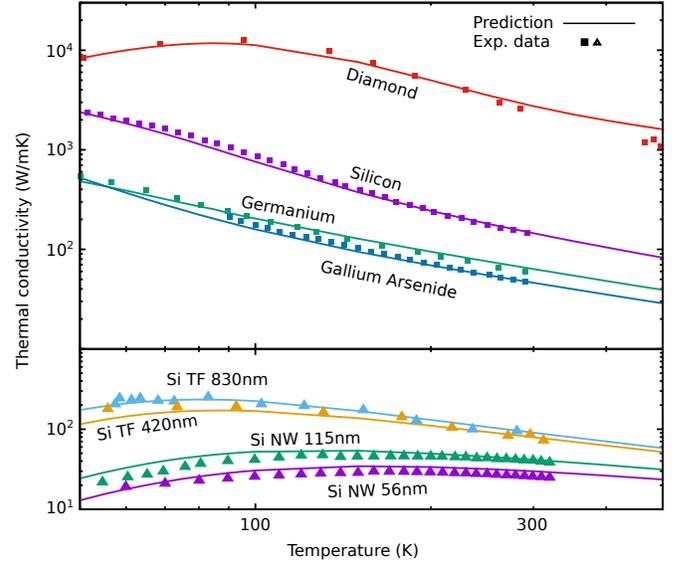}
\caption{Top: Prediction of thermal conductivity for bulk Si, Ge, Diamond and GaAs from Eq. (\ref{KCM}) using {\it ab initio} scattering rates. Bottom: Predictions of thermal conductivity for thin films and nanowires. Points represent experimental data, lines the theoretical predictions.}\label{fig_bulk_kappa}
\end{figure}

In Fig.~\ref{fig_bulk_kappa} top we plot the calculated thermal conductivity from {\it first principles} with KCM and compare them to experimental measurements for bulk Si, Ge, diamond and GaAs samples. The remarkable agreement of predictions and experiments without any adjustable parameter shows that the model is set on solid grounds. Similar results for bulk samples have also been obtained using a different approach based on an iterative solution of the BTE \cite{Broido2007}. In the latter model, the effect of the normal scattering process is included through the iterative process, whereas in KCM this is determined by $\Sigma$. Fig. \ref{fig_bulk_kappa} bottom shows the KCM predictions for nanowires and films. As no error bars are provided by the experimental data in the whole temperature range, the reduction in the thermal conductivity for the nanosamples included through $ \tau_B $ and F can be considered as a good approach. Comparison to experimental data for silicon nanowires and films using a parameter free approach has not been published yet. In this line, pure kinetic models can provide good fits with data; however, they are not fully satisfactory because the parameters of the intrinsic relaxation times used at small scales do not agree to the bulk ones~\cite{Martin2009b}.More efforts are needed to fully understand the heat transport at the nanoscale.

\begin{figure}[h!]
\includegraphics[width=\columnwidth]{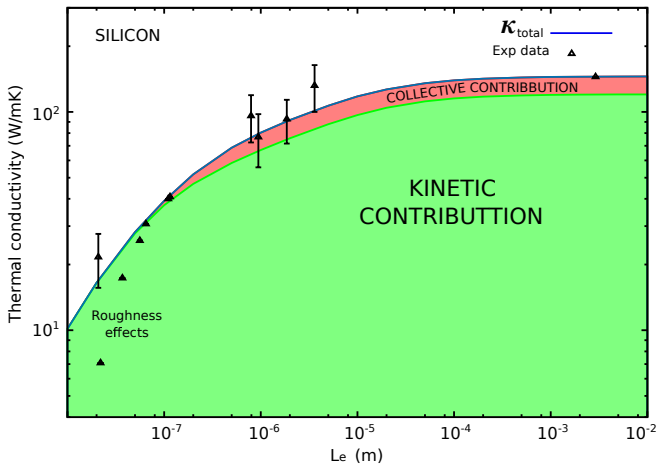}
\caption{Kinetic, collective and total thermal conductivity for silicon wires and films as a function of the effective length at T=300K. Experimental data are taken from \cite{Li2003} for nanowires, from \cite{Asheghi1997} for films and from \cite{Ozhogin1996} for bulk. $L_{\rm eff}=d_{\rm wire}$ and $L_{\rm eff}=2.25h_{\rm film}$(see \cite{DeTomas2014}).}\label{size_factors}
\end{figure}

\begin{figure*}
\includegraphics[width=0.85\textwidth]{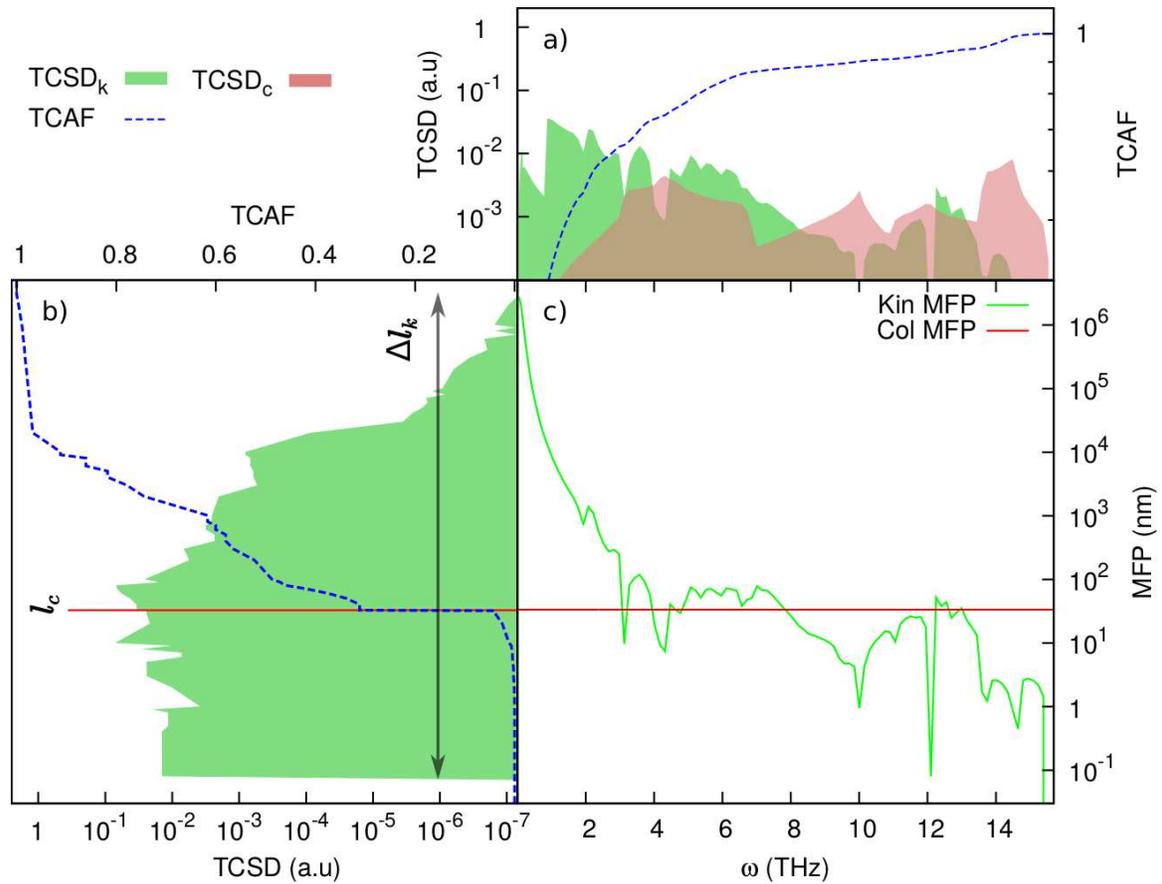}
\caption{Thermal conductivity spectral distribution (TCSD) in terms of (a) frequency and (b) mean free path for silicon at T=300K. Filled curves in (a) and (b) are the kinetic and collective contributions to TCSD. Blue line represents the thermal conductivity acumulation function (TCAF), that is, the integral of TCSD. Plot (c) represents the kinetic and collective mean free paths ($\ell_k$ and $\ell_{c}$) in terms of frequency.}\label{fig_MFP}
\end{figure*}

Fig. \ref{size_factors} presents KCM calculations for five samples going from bulk to 20 nm nanowires, showing a remarkable agreement with experimental data. Notice that KCM gives good predictions without any fitting parameter for nanowires as small as 56nm as no roughness effects are needed. The two smallest wires have been reported to exhibit roughness~\citep{Hochbaum2008,Martin2009b}, consequently the KCM prediction without roughness effects overestimate the thermal conductivity. The comparison of the KCM predictions with those of a pure kinetic model sheds light on why matching bulk and small sample experiments has been elusive in the latter models. The green zone in Fig. \ref{size_factors} displays the kinetic contributtion to thermal conductivity contribution, namely $\kappa_k$. The difference between the blue line and the green zone is the collective contributtion (red zone). While the agreement with data is good for small diameters (where the red zone vanishes), for bulk (diameter=3mm) the collective contributtion is not negligible. It is the right combination of kinetic and collective regimes as expressed by Eq. (\ref{KCM}) that yields accurate predictions in the whole size range. The convergence of $\kappa_{tot}$ to $\kappa_{k}$ for small samples can be explained as follows. As size is reduced, the rate of boundary collisions increases while normal scattering rates do not change. As a result, parameter $\Sigma$ becomes smaller and so does the weight of the collective contribution.

Let us note that the differences between kinetic and collective contributions are important to interpret the phonon spectrum. Fig.~\ref{fig_MFP} displays the thermal conductivity spectrum distribution (TCSD) for silicon in terms of (a) frequency and (b) mean free path (MFP). The filled green and red curves show the kinetic and collective contributions. The dashed lines show the thermal conductivity accumulation function (TCAF), that is, the integral of TCSD. In terms of frequency [Fig.~\ref{fig_MFP}(a)], both contributions span the whole range of the spectra, in contrast in terms of the MFP [Fig.~\ref{fig_MFP}(b)], they have significant differences. In Fig.~\ref{fig_MFP}(c) the kinetic and collective mean free paths $\ell_k(\omega)=v_k(\omega)\tau_k(\omega)$ and $\ell_c=v_c\tau_c$ are represented in terms of frequency, where $v_k(\omega)$ is the mode velocity depending on frequency $\omega$ and 
\begin{equation}
v_c=\frac{\int v_k(\omega) C_{v}(\omega)DOS(\omega) d\omega}{\int C(\omega) DOS(\omega)d\omega}
\end{equation}
is the average collective velocity (independent of mode); $C(\omega)$ is the specific heat per mode. Notice that while the kinetic MFP is different for all the modes, the collective MFP is the same for all of them. Consequently, the collective TCSD in Fig.~\ref{fig_MFP}(b) is reduced to a delta function at a single point ($\ell_{c}$) of the spectrum, while the kinetic TCSD spans an extended region of MFP ($\Delta \ell_{k}$). As a consequence, the TCAF rises in a single step at the point where collective modes add their contribution (blue line in Fig.~\ref{fig_MFP}.b). This raise seems to appear in previous works~\citep{Vermeersch2015,Cuffe2015,Zeng2015}, however the use of a pure kinetic model did not allow to identify it with a collective regime. As the MFP spectra depends on the model used, it is expected to find different behaviours depending on the model. 

Notice that the differences between the kinetic and the collective distributions have also an impact on the reconstruction of the TCSD, as one can obtain different results if kinetic and collective contributions are located in their corresponding MFP, that is $\ell_k(\omega)$ and $\ell_c$, or if total MFP is used $\ell_{tot}(\omega)=(1-\Sigma)\ell_{k}(\omega)+\Sigma\ell_{c}$ instead. This might be the reason for the lack of abrupt changes in the obtained reconstructions using kinetic approaches in recent ultra-fast heating experiments.~\cite{Regner2013,Johnson2013,Wilson2014,Hu2015,Hoogeboom-Pot2015}

Note also that the differences between the kinetic and the collective MFP distributions [Fig.~\ref{fig_MFP}.b)] can also provide an explanation for the deviations from Fourier law observed in some experiments. The presence of a large range of scales $\Delta \ell_{k}$ would explain the appearance of superdiffusivity in alloys as recently proposed~\citep{Vermeersch2015, Vermeersch2015a}, while the single scale collective regime $\ell_{c}$ leads to different behaviors like Poiseuille flow or second sound in materials where normal scattering is dominant~\cite{Guyer1966,Lee2015,Cepellotti2015}.



\begin{figure} [h!]
\includegraphics[width=\columnwidth]{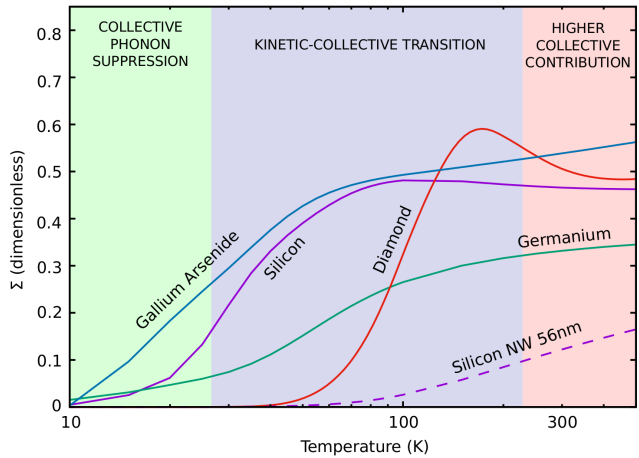}
\caption{Switching factor $\Sigma$ for bulk Si, Ge, diamond, GaAs and a 56~nm SiNW as a function of temperature. One observes a transition from a pure kinetic regime at small temperatures (small $ \Sigma$) to a combination of kinetic and collective transport at higher temperatures.}\label{fig:sigma}
\end{figure}

Finally, the relative weight of the kinetic and collective terms also depends on temperature; therefore an impact on the transport regime will be expected. Fig. ~\ref{fig:sigma} shows the value of the switching factor $\Sigma$ for different materials as a function of temperature. It can be observed that the collective contribution becomes significant in all cases as the temperature raises, achieving a constant value at high temperatures. At low temperatures, as $\tau_{n}(T)$ increases and in $\tau_{k}$ the boundary and impurity terms are temperature independent, from the $\Sigma$ definition (~\ref{eq_sigma}) it is clear that the collective contribution vanishes. This is key to interpret experiments where different temperatures are used~\cite{Minnich2011,Zeng2015}. Notice that collective effects are important even at temperatures as low as 150K, so we can only expect pure kinetic models to be valid at very low temperatures. This information can be combined with Fig. \ref{fig_MFP}(b) to see that as the temperature raises, the TCAF distribution in terms of the MFP will experience a gradual change from a kinetic distribution, smoothly spanned over $\Delta\ell_{k}$, to a more collective distribution, with a steeper slope around $\ell_{c}$.

In conclusion, KCM offers a unifying framework to understand size and temperature effects in samples where normal scattering plays a significant role. Its predictions using \textit{first principles} are in excellent agreement with experimental values for all the materials studied, without free parameters. Our results stress the importance of determining the contribution of collective transport to interpret properly the results of an experimental setup. The precise separation of kinetic and collective contributions that the model supplies is expected to shed light on the behaviour of thermal conductivity in high gradient temperature and ultra-fast experiments like pump-probe and thermoreflectance, where dynamic effects on the phonon distribution can be produced, leading to the possibility of identifying the appearance of memory and non-local effects. 

We acknowledge the financial support of the Spanish Ministry of Economy and Competitiveness under Grant Consolider nanoTHERM CSD2010-00044, TEC2015-67462-C2-2-R (MINECO/FEDER), TEC2015-67462-C2-1-R (MINECO/FEDER)

\bibliographystyle{apsrev}

\end{document}